\documentstyle[runningheads,fleqn,psfig]{svmult}

\begin{document}
\title*{Hot Dark Matter in Cosmology}
\toctitle{Hot Dark Matter in Cosmology}
%
%
\titlerunning{Hot Dark Matter}
%
\author{Joel R. Primack
\and Michael A. K. Gross}


\institute{Physics Department, University of California, Santa Cruz, CA
95064 USA}


%
\authorrunning{Primack and Gross}
%
%

\maketitle              

\def\ltsima{$\; \buildrel < \over \sim \;$}
\def\lsim{\lower.5ex\hbox{\ltsima}}
\def\gtsima{$\; \buildrel > \over \sim \;$}
\def\gsim{\lower.5ex\hbox{\gtsima}}
\def\lcdm{$\Lambda$CDM}
\def\lchdm{$\Lambda$CHDM}
\def\hmpc{h^{-1} \, {\rm Mpc}}
\def\epem{e^+ e^-}


\section{Historical Summary}

Cosmological dark matter in the form of neutrinos with masses of up to
a few electron volts is known as hot dark matter.  In 1979-83, this
appeared to be perhaps the most plausible dark matter candidate.  Such
HDM models of cosmological structure formation led to a top-down
formation scenario, in which superclusters of galaxies are the first
objects to form, with galaxies and clusters forming through a process
of fragmentation.  Such models were abandoned when it was realized
that if galaxies form sufficiently early to agree with observations,
their distribution would be much more inhomogeneous than it is
observed to be.  Since 1984, the most successful structure formation
models have been those in which most of the mass in the universe is in
the form of cold dark matter (CDM).  But mixed models with both cold
and hot dark matter (CHDM) were also proposed in 1984, although not
investigated in detail until the early 1990s.

The recent atmospheric neutrino data from Super-Kamiokande provide
strong evidence of neutrino oscillations and therefore of non-zero
neutrino mass.  These data imply a lower limit on the HDM (i.e., light
neutrino) contribution to the cosmological density $\Omega_\nu \gsim
0.001$ --- almost as much as that of all the stars in the centers of
galaxies --- and permit higher $\Omega_\nu$.  The ``standard''
COBE-normalized critical-matter-density (i.e., $\Omega_m=1$) CDM model
has too much power on small scales.  It was discovered in 1992-95 that
CDM with the addition of neutrinos with total mass of about 5 eV,
corresponding to $\Omega_\nu \approx 0.2$, results in a much improved
fit to data on the nearby galaxy and cluster distribution.  Indeed,
the resulting Cold + Hot Dark Matter (CHDM) cosmological model is
arguably the most successful $\Omega_m=1$ model for structure
formation \cite{Primack95,Primack96,GawiserSilk98,Gross98}.

However, other recent data have begun to make a convincing case for
$0.3 \lsim \Omega_m \lsim 0.5$.  In light of all these new data,
several authors have considered whether cosmology still provides
evidence favoring neutrino mass of a few eV in flat models with
cosmological constant $\Omega_\Lambda = 1 - \Omega_m$. The conclusion
is that the possible improvement of the low-$\Omega_m$ flat (\lcdm)
cosmological models with the addition of light neutrinos appears to be
rather limited, but that \lchdm\ models with $\Omega_\nu \lsim 0.1$
may be consistent with presently available data.  Data expected soon
may permit detection of such a hot dark matter contribution, or
alternatively provide stronger upper limits on $\Omega_\nu$ and
neutrino masses.

\section{Hot, Warm, and Cold Dark Matter}

Hot DM refers to particles, such as neutrinos, that were moving at
nearly the speed of light at redshift $z\sim10^6$ (or time $t\sim 1$
yr), when the temperature $T\sim 3\times10^2$ eV and the cosmic
horizon first encompassed $10^{12}M_\odot$, the amount of dark matter
contained in the halo of a large galaxy like the Milky Way.  Hot DM
particles must also be still in thermal equilibrium after the last
phase transition in the hot early universe, the QCD confinement
transition, which presumably took place at $T_{\rm QCD} \approx 10^2$
MeV. Hot DM particles have a cosmological number density roughly
comparable to that of the microwave background photons, which (as we
will see shortly) implies an upper bound to their mass of a few tens
of eV. This then implies that free streaming of these relativistic
particles destroys any fluctuations smaller than supercluster size,
$\sim10^{15}M_\odot$.

The ``hot,'' ``warm,'' ``cold'' DM terminology was introduced in 1983
\cite{Bond83,PB83}.  Warm DM particles interact much more weakly than
neutrinos. They decouple (i.e., their mean free path first exceeds the
horizon size) at $T \gg T_{\rm QCD}$, and are not heated by the
subsequent annihilation of hadronic species.  Consequently their
number density is roughly an order of magnitude lower, and their mass
an order of magnitude higher, than hot DM particles. Fluctuations
corresponding to sufficiently large galaxy halos, $\gsim
10^{11}M_\odot$, could then survive free streaming. In theories of
local supersymmetry broken at $\sim 10^6$ GeV, gravitinos could be DM
of the warm variety \cite{PagelsP82,BlumPP82,BondST82}.  Other warm
dark matter candidates are also possible, of course, such as
right-handed neutrinos \cite{OliveT82}.  Warm DM does not fit the
observations if $\Omega_m=1$ \cite{WDM96}, but for low $\Omega_m$ some
have suggested that it may be worth reconsidering, to avoid some
possible problems of Cold DM \cite{SLDolgov,Hogan}.  However, the
cutoff in the power spectrum $P(k)$ at large $k$ implied by WDM will
also inhibit the formation of small dark matter halos at high
redshift.  But such small halos are presumably where the first stars
form, which produce metals rather uniformly throughout the early
universe as indicated by observations of the Lyman $\alpha$ forest
(neutral hydrogen clouds seen in absorption in quasar spectra).

Cold DM consists of particles for which free streaming is of no
cosmological importance. Two different sorts of cold DM consisting of
elementary particles have been proposed, heavy thermal remnants of
annihilation such as supersymmetric neutralinos, and a cold Bose
condensate such as axions.  A universe where the matter is mostly cold
DM and there is a large cosmological constant looks very much like the
one astronomers actually observe, and this low-$\Omega_m$ $\Lambda$CDM
model \cite{BFPR} is the current favorite model for structure
formation in the universe \cite{Bahcall99,Turner99,P99}.

\section{Galaxy Formation with Hot DM}

The standard hot DM candidate is massive neutrinos
\cite{GZ66,MS71,CMc72,SM76}, although other, more exotic theoretical
possibilities have been suggested, such as a ``majoron'' of nonzero
mass which is lighter than the lightest neutrino species, and into
which all neutrinos decay.  Neutrinos appeared to be an attractive DM
candidate because of the measurement of an electron neutrino mass of
about 30 eV in 1980 \cite{Lyu80}.  This coincided with the improving
CMB limits on the primordial fluctuation amplitude, which forced
Zel'dovich and other theorists to abandon the idea that all the dark
matter could be made of ordinary baryonic matter.  The version of HDM
that they worked out in detail, with adiabatic Gaussian primordial
fluctuations, became the prototype for the subsequent $\Omega_m=1$ CDM
theory.

\subsection{Mass Constraints}

Direct measurements of neutrino masses have given only upper limits
(see also the chapter by Robertson and Wilkerson).  A secure upper
limit on the electron neutrino mass is roughly 15 eV.  The Particle
Data Group \cite{Databook} notes that a more precise limit cannot be
given since unexplained effects have resulted in significantly
negative measurements of $m(\nu_e)^2$ in tritium beta decay
experiments.  However, this problem is at least partially resolved,
and the latest experimental upper limits on the electron neutrino mass
are 2.8 eV from the Mainz \cite{Mainz} and 2.5 eV from the Troitsk
\cite{Troitsk} tritium beta decay experiments (both 95\% C.L.).  There
is an upper limit on an effective Majorana neutrino mass of $\sim 1$
eV from neutrinoless double beta decay experiments \cite{Bilenky}.
The upper limits from accelerator experiments on the masses of the
other neutrinos are $m(\nu_\mu)< 0.17$ MeV (90\% CL) and $m(\nu_\tau)<
18$ MeV (95\% CL) \cite{Databook,NuIndustry}, but since stable
neutrinos with such large masses would certainly ``overclose the
universe'' (i.e., contribute such a large cosmological density that
the universe could never have attained its present age), cosmology
implies a much lower upper limit on these neutrino masses.

Before going further, it will be necessary to discuss the thermal
history of neutrinos in the standard hot big bang cosmology in order
to derive the corresponding constraints on their mass.
Left-handed neutrinos of mass $\leq$ 1 MeV remain in thermal
equilibrium until the temperature drops to $T_{\nu d}$, at which point
their mean free path first exceeds the horizon size and they
essentially cease interacting thereafter, except gravitationally
\cite{Weinberg72}.  Their mean free path is, in natural units $(\hbar
= c = 1)$, $\lambda_\nu \sim [\sigma_\nu n_{e^\pm}]^{-1} \sim [(G^2_F
T^2) (T^3)]^{-1}$, where $G_F \approx 10^{-5}$ GeV$^{-2}$ is the Fermi
constant that measures the strength of the weak interactions.  The
horizon size is $\lambda_h \sim (G \rho)^{-1/2} \sim M_{P\ell}T^{-2}$,
where the Planck mass $M_{P\ell} \equiv G^{-1/2} = 1.22 \times
10^{19}$ GeV.  Thus $\lambda_h/\lambda_\nu \sim (T/T_{\nu d})^3$, with
the neutrino decoupling temperature
\begin{equation}
T_{\nu d} \sim M_{P\ell}^{-1/3} G_F ^{-2/3} \sim 1\, {\rm MeV} \ .
\end{equation}

After $T$ drops below ${1\over2}$ MeV, $e^+e^-$ annihilation ceases
to be balanced by pair creation, and the entropy of the $\epem$
pairs heats the photons.
Above 1 MeV, the number density $n_{\nu i}$ of each left-handed
neutrino species and its right-handed antiparticle is
equal to that of the photons, $n_\gamma$,
times the factor 3/4 from Fermi versus Bose statistics.  But then
$\epem$ annihilation increases the photon number density relative
to that of the neutrinos by a factor of 11/4.(\footnote{In
the argument giving the 11/4 factor,
the key ingredient is that the entropy in
interacting particles in a comoving volume $S_I$ is conserved
during ordinary Hubble expansion, even during a process such
as electron-positron annihilation, so long as it occurs in
equilibrium.  That is, $S_I=g_I(T) N_\gamma(T)={\rm constant}$, where
$N_\gamma=n_\gamma V$ is the number of photons in a given
comoving volume $V$, and $g_I=(g_B+{7\over8}g_F)_I$ is the
effective number of helicity states in interacting particles
(with the factor of ${7\over8}$ reflecting the difference in
energy density for fermions versus bosons).  Just above the
temperature of electron-positron annihilation,
$g_I=g_\gamma+{7\over8}\times
g_e=2+{7\over8}\times4={11\over2}$; while below it,
$g_I=g_\gamma=2$.  Thus, as a result of the entropy of the
electrons and positrons being dumped into the photon gas at
annihilation, the photon number density is thereafter
increased relative to that of the neutrinos by a factor of
11/4.})  As a result, the neutrino temperature $T_{\nu,0} =
(4/11)^{1/3} T_{\gamma,0}$.
Thus today, for each species,
\begin{equation}
 n_{\nu,0} = {3\over 4}\cdot{4\over 11} \, n_{\gamma,0}
= 109\, \theta^3 {\rm cm}^{-3}\ ,
\end{equation}
where $\theta \equiv (T_0/2.7 {\rm K})$.
With the cosmic background radiation temperature $T_0=2.728 \pm 0.004$ K
measured by the FIRAS instrument on the COBE satellite \cite{CMBT},
$T_{\nu,0} = 1.947$ K and $n_{\nu,0}=112\,{\rm cm}^{-3}$.

Since the present cosmological matter density is
\begin{equation}
 \bar\rho_m = \Omega\rho_c = 10.54\, \Omega_m h^2 {\rm keV}\,{\rm cm}^{-3} \ ,
\end{equation}
it follows that
\begin{equation} \sum_i m_{\nu i} < \bar\rho_m/n_{\nu,o}
 \leq 96\, \Omega_m h^2 \theta^{-3} {\rm eV}
 \approx 93\, \Omega_m h^2 {\rm eV}\ ,
\label{eq:nubound}
\end{equation}
where the sum runs over all neutrino species with
$M_{\nu i} \leq$ 1 MeV. (Heavier neutrinos will be discussed in the
next paragraph.)
Observational data imply that $\Omega_m h^2 \approx 0.1-0.3$, since
$\Omega_m \approx 0.3-0.5$ and $h \approx 0.65 \pm 0.1$ \cite{P99}.
Thus if all the dark matter were light neutrinos, the sum of their
masses would be $\approx 9-28$ eV.

In deriving eq. (\ref{eq:nubound}), we have been assuming that all the
neutrino species are light enough to still be relativistic
at decoupling, i.e. lighter than an MeV.  The bound (\ref{eq:nubound})
shows that they must then be much lighter than that.  In the
alternative case that a neutrino species is nonrelativistic
at decoupling, it has been shown
\cite{LeeW77,Vys77,Hut77,SatoK77,Gunn78}
that its mass must then exceed several GeV, which is not
true of the known neutrinos ($\nu_e$, $\nu_\mu$, and
$\nu_\tau$).  (One might at first think that the Boltzmann factor
would sufficiently suppress the number density of
neutrinos weighing a few tens of MeV to allow compatibility with
the present density of the universe.  It is the fact that they
``freeze out'' of equilibrium well before the temperature drops to
their mass that leads to the higher mass limit.)
We have also been assuming that the neutrino
chemical potential is negligible, i.e. that
$\left | n_\nu-n_{\bar\nu}\right | \ll n_\gamma$.
This is very plausible,
since the net baryon number density $(n_b-n_{\bar b}) \lsim
10^{-9} n_\gamma$, and big bang nucleosynthesis restricts the allowed
parameters \cite{KangSteigman} (see also the chapter by Fuller).

\subsection{Phase Space Constraint}

We have just seen that light neutrinos must satisfy an upper bound on
the sum of their masses.  But now we will discuss a lower bound on
neutrino mass that arises because they must be rather massive to form
the dark matter in galaxies, since their phase space density is
limited by the Pauli exclusion principle.  A slightly stronger bound
follows from the fact that they were not degenerate in the early
universe.

The phase space constraint \cite{TremainGunn79}
follows from Jeans's theorem in classical mechanics to the effect that
the maximum 6-dimensional phase space density cannot increase as a
system of collisionless
particles evolves.  At early times, before density
inhomogenitites become nonlinear, the neutrino phase space
density is given by the Fermi-Dirac distribution
\begin{equation}
n_\nu(p)={g_\nu \over h^3}
   \left[1+\exp({pc \over kT_\nu(z)})\right]^{-1},
\end{equation}
where here $h$ is Planck's constant and
$g_\nu=2$ for each species of left-handed $\nu$ plus right-handed 
$\bar \nu$.
Since momentum and temperature both scale as redshift $z$ as the
universe expands, this distribution remains valid after neutrinos
drop out of thermal equilibrium at $\sim 1$ MeV, and even into the
nonrelativistic regime $T_\nu < m_\nu$ \cite{Weinberg72}.
The standard version of the
phase space constraint follows from demanding that
the central phase space density
$9[2 (2\pi)^{5/2} G r_c^2 \sigma m_\nu^4]^{-1}$ of the DM halo,
assumed to be
an isothermal sphere of core radius $r_c$ and one-dimensional
velocity dispersion $\sigma$, not exceed the maximum value of the
initial phase space density $n_\nu(0)= g_\nu/{2 h^3}$.
The result is
\begin{equation}
m_\nu > (120\,{\rm eV})
  \left({100\,{\rm km\,s}^{-1}\over \sigma}\right)^{1/4}
  \left({1\,{\rm kpc}\over r_c}\right)^{1/2}
  \left({g_\nu \over 2}\right)^{-1/4}.
\end{equation}

The strongest lower limits on $m_\nu$ follow from applying this to the
smallest galaxies.
Both theoretical arguments regarding the dwarf spheroidal
($dS$) satellite galaxies of the Milky Way \cite{FaberLin83}
and data on Draco, Carina, and Ursa Minor made it clear some time ago that
dark matter dominates the gravitational potential of
these $dS$ galaxies, and the case has only strengthened with
time \cite{PryorKormendy90}.
The phase space constraint
then sets a lower limit \cite{LinFaber83}
$m_\nu >$ 500 eV, which is completely incompatible with the
cosmological constraint eq. (\ref{eq:nubound}). However, this
argument only excludes neutrinos as the DM in certain small
galaxies; it remains possible that the DM in these galaxies
is (say) baryonic, while that in larger galaxies such as our
own is (at least partly) light neutrinos.
A more conservative phase space constraint was obtained for the Draco
and Ursa Minor dwarf spheroidals \cite{GerhardSpergel92}, but the
authors concluded that neutrinos consistent with the cosmological upper
bound on $m_\nu$ cannot be the DM in these galaxies.  A similar
analysis applied to the gas-rich low-rotation-velocity dwarf irregular
galaxy DDO 154 \cite{SWGott88} gave a limit $m_\nu >94$ eV, again
inconsistent with the cosmological upper bound.

\subsection{Free Streaming}

The most salient feature of hot DM is the erasure of small
fluctuations by free streaming. Thus even collisionless particles
effectively exhibit a Jeans mass.  It is easy to see that the
minimum mass of a surviving fluctuation is of order
$M^3_{P\ell}/m^2_\nu$ \cite{BKN80,BES80}.
Let us suppose that some process in the very early
universe --- for example, thermal fluctuations subsequently
vastly inflated in the inflationary scenario ---
gave rise to adiabatic fluctuations on all scales.  In adiabatic
fluctuations, all the components --- radiation and matter ---
fluctuate together.
Neutrinos of nonzero mass $m_\nu$ stream relativistically
from decoupling until the temperature drops to $T \sim m_\nu$,
during which time they traverse a distance $d_\nu =
R_H(T = m_\nu) \sim M_{P\ell}\, m_\nu^{-2}$. In order to
survive this free streaming, a neutrino fluctuation must be
larger in linear dimension than $d_\nu$. Correspondingly,
the minimum mass in neutrinos of a surviving fluctuation is
$M_{J,\nu} \sim d^3_\nu m_\nu n_\nu (T = m_\nu) \sim d^3_\nu
m^4_\nu \sim M^3_{P\ell}\, m_\nu^{-2}$. By analogy with Jeans's
calculation of the minimum mass of an ordinary fluid
perturbation for which gravity can overcome pressure, this
is referred to as the (free-streaming) Jeans mass.

A more careful calculation \cite{BES80,BS84} gives
\begin{equation}
 d_\nu = 41 (m_\nu/30\,{\rm eV})^{-1} (1+z)^{-1} {\rm Mpc}\ ,
 \end{equation}
that is, $d_\nu = 41 (m_\nu/30{\rm eV})^{-1}$ Mpc in comoving
coordinates, and correspondingly
\begin{equation} M_{J,\nu} = 1.77\, M^3_{P\ell}\,m^{-2}_\nu
 = 3.2 \times 10^{15}(m_\nu/30\, {\rm eV})^{-2}M_\odot \ ,
\label{eq:MsubJnu}
\end{equation}
which is the mass scale of superclusters. Objects of this
size are the first to form in a $\nu$-dominated universe,
and smaller scale structures such as galaxies can form only
after the initial collapse of supercluster-size
fluctuations.

When a fluctuation of total
mass $\sim 10^{15} M_\odot$ enters the horizon at $z \sim
10^4$, the density contrast $\delta_{\rm RB}$ of the
radiation plus baryons ceases growing and instead starts
oscillating as an acoustic wave, while that of
the massive neutrinos $\delta_\nu$ continues to grow
linearly with the scale factor $R = (1+z)^{-1}$ since the
Compton drag that prevents growth of $\delta_{\rm RB}$ does
not affect the neutrinos.  By recombination, at $z_r
\sim 10^3$, $\delta_{\rm RB}/\delta_\nu \lsim 10^{-1}$, with
possible additional suppression of $\delta_{\rm RB}$ by Silk
damping. Thus the hot DM scheme with adiabatic primordial fluctuations
predicts small-angle fluctuations in the microwave
background radiation that are lower than in the
adiabatic baryonic cosmology, which was one of the reasons HDM
appealed to Zel'dovich and other theorists.  Similar considerations
apply in the warm and cold DM schemes. However, as we will discuss in
a moment, the HDM top-down 
sequence of cosmogony is wrong, and with the COBE
normalization hardly any structure would form by the present.

In numerical simulations of dissipationless gravitational
clustering starting with a fluctuation spectrum
appropriately peaked at $\lambda \sim d_\nu$ (reflecting
damping by free streaming below that size and less time for
growth of the fluctuation amplitude above it),
the regions of high density form a network of
filaments, with the highest densities occurring at the
intersections and with voids in between
\cite{Melott83,CMelott83,KS83,FWD83}.  
The similarity of these
features to those seen in observations was cited as evidence
in favor of HDM \cite{ZelES82}.

\subsection{Problems with $\nu$ DM}

A number of potential problems with the neutrino dominated
universe had emerged by about 1983, however.
\begin{itemize}
\item
From studies both of nonlinear clustering
\cite{FWD83,DekelA84}
(comoving length scale $\lambda \lsim 10$ Mpc) and of
streaming velocities \cite{Kaiser83}
in the linear regime ($\lambda > 10$ Mpc), it follows that
supercluster collapse must have occurred recently:
$z_{sc} \leq 0.5$ is indicated and in any case $z_{sc} < 2$
\cite{FWD83}.
However, the best limits on galaxy ages coming from globular
clusters and other stellar populations indicated that galaxy
formation took place before $z \approx 3$.
Moreover, if quasars are associated with galaxies, as is
suggested by the detection of galactic luminosity around
nearby quasars and the apparent association of more distant
quasars with galaxy clusters, the abundance of quasars at $z > 2$
was also inconsistent with the ``top-down'' neutrino
dominated scheme in which superclusters form first:  $z_{sc}
> z_{galaxies}$.
\item
Numerical simulations of the nonlinear ``pancake'' collapse
taking into account dissipation of the baryonic matter showed
that at least 85\%\ of the baryons are so heated by the
associated shock that they remain unable to
condense, attract neutrino halos, and eventually form
galaxies \cite{Bond83,SSMM83}.
This was a problem for the hot DM scheme for two
reasons.  With the primordial nucleosynthesis constraint
$\Omega_b\lsim0.1$, there would be difficulty having enough
baryonic matter condense to form the luminosity that we
actually observe.  And, where are the X-rays from the
shock-heated pancakes \cite{White86}?
\item
The neutrino picture predicts \cite{BSW83}
that there should be a factor of
$\sim5$ increase in $M/M_b$ between large galaxies
($M \sim 10^{12} M_\odot$) and large clusters ($M
\geq 10^{14}M_\odot$), since the larger clusters,
with their higher escape velocities, are able to trap a
considerably larger fraction of the neutrinos.  Although
there is some indication that the mass-to-light ratio
$M/L$ increases with $M$, the ratio of total to
luminous mass $M/M_{lum}$ is probably a better indicator of
the value of $M/M_b$, and
it is roughly the same for galaxies with large halos
and for rich clusters.
\end{itemize}

These problems, while serious, would perhaps not have been fatal
for the hot DM scheme.  But an even more
serious problem for HDM arose from the low
amplitude of the CMB fluctuations detected by the COBE satellite,
$(\Delta T/T)_{rms}=(1.1\pm0.2)\times 10^{-5}$ smoothed on an angular
scale of about $10^\circ$ \cite{Smoot92}. Although HDM and CDM both have the
Zel'dovich spectrum shape ($P(k) \propto k$) in the long-wavelength
limit, because of the
free-streaming cutoff the amplitude of the HDM spectrum must be
considerably higher in order to form any structure by the present.
With the COBE normalization, the HDM spectrum is
only beginning to reach nonlinearity at the present epoch.

Thus the evidence against standard hot DM is convincing. At very
least, it indicates that structure formation in a neutrino-dominated
universe must be rather more complicated than in the standard
inflationary picture.

The main alternative that has been considered is cosmic strings plus
hot dark matter.  Because the strings would continue to seed structure
up until the present, and because these seeds are in the nature of
rather localized fluctuations, hot DM would probably work better with
string seeds than cold DM.  However, strings and other cosmic defect
models are now essentially ruled out \cite{Pen97,Albrecht99} 
because they predict that the
cosmic microwave background would have an angular power spectrum
without the pronounced (doppler/acoustic/Sakharov) peak at angular
wavenumber $l \sim 220$ that now appears to be clearly indicated by the
data, along with secondary peaks at higher $l$.

\section{Cold plus Hot Dark Matter and Structure Formation: $\Omega_m=1$}

Even if most of the dark matter is of the cold variety, a little hot
dark matter can have a dramatic effect on the predicted distribution of
galaxies.  In the early universe, the free streaming of the fast-moving
neutrinos washes out any inhomogeneities in their spatial distribution
on the scales that will later become galaxies.  If these neutrinos are
a significant fraction of the total mass of the universe, then although
the density inhomogeneities will be preserved in the cold dark matter,
their growth rates will be slowed.  As a result, the amplitude of the
galaxy-scale inhomogeneities today is less with a little hot dark
matter than if the dark matter is only cold.  (With the tilt $n$ of
the primordial spectrum $P_p(k) = A k^n$ fixed --- which as we
discuss below is not necessarily reasonable --- the fractional
reduction in the power on small scales is $\Delta P/P \approx 8
\Omega_\nu / \Omega_m$ \cite{Hu98}.  See Fig. 1 for examples of how
the power spectrum $P(k)$ is affected by the addition of hot dark matter in
$\Omega_m = 0.4$ flat cosmologies.)  Since the main problem with
$\Omega_m=1$ cosmologies containing only cold dark matter is that the
amplitude of the galaxy-scale inhomogeneities is too large compared to
those on larger scales, the presence of a little hot dark matter
appeared to be possibly just what was needed.  And, as was mentioned at
the outset, a CHDM model with $\Omega_m=1$, $\Omega_\nu=0.2$, and
Hubble parameter $h=0.5$ is perhaps the best fit to the galaxy
distribution in the nearby universe of any cosmological model.  The
effects of the relatively small amount of hot dark matter in a CHDM
model on the distribution of matter compared to a purely CDM model are
shown graphically in \cite{Brodbeck}; cf. also \cite{Ma99}.  As
expected, within galaxy halos the distribution of cold and hot
particles is similar.  But the hot particles are more widely
distributed on larger scales, and the hot/cold ratio is significantly
enhanced in low-density regions.

The first step in working out the theory of structure formation is to
use linear perturbation theory, which is valid since cosmic microwave
background measurements show that density fluctuations are small at the
redshift of recombination, $z_r \sim 10^3$.  The most extensive early
calculations of this sort were carried out by Holtzman
\cite{Holtz89,HoltzKlyp98}, who concluded that the most promising 
cosmological models were CHDM and $\Lambda$CDM \cite{HoltzPri}.
The most efficient method of
computing the linear evolution of fluctuations now is that used in the
CMBFAST code \cite{Seljak}.  An alternative Monte Carlo treatment of
the evolution of neutrino density fluctuations was given by
\cite{MaBert95}, but the differences from the usual treatment appear to
be small.  Detailed analytic results have been given by
\cite{Ma96,EisensteinHu} and reviewed in \cite{Ma99}.  But the key
point can be understood simply: there is less structure in CHDM models
on small scales because the growth rate of cold dark matter
fluctuations is reduced on the scales where free streaming has wiped
out neutrino fluctuations.  Let us define the fluctuation growth rate
$f$ by
\begin{equation}
f(k) \equiv {{d \log \delta(k)} \over {d \log a}} \ ,
\end{equation}
where $\delta(k)$ is the amplitude of the fluctuations of wave number
$k=2\pi/\lambda$ in cold dark matter, and as usual $a=1/(1+z)$ is the
scale factor.  For $\Omega_m=1$ CDM fluctuations, the growth rate
$f=1$.  This is also true for fluctuations in CHDM, for $k$
sufficiently small that free-streaming has not significantly
decreased the amplitude of neutrino fluctuations.  However, in the
opposite limit $k \longrightarrow \infty$ \cite{BES80,Ma99},
\begin{equation}
f_\infty = (\sqrt{1+24\Omega_c} - 1)/4 \approx \Omega_c^{0.6} \ ,
\end{equation}
assuming that $\Omega_c + \Omega_\nu = 1$.  For example, for
$\Omega_\nu = 0.2$, $f_\infty=0.87$.  Even though the growth rate
is only a little lower for these large-$k$ (i.e., short-wavelength)
modes, the result is that their amplitude is decreased substantially
compared to longer-wavelength modes.

The next step in determining the implications for structure formation
is to work out the effects on nonlinear scales using N-body
simulations.  This is harder for Cold+Hot models than for CDM because
the higher velocities of the neutrinos require more particles to
adequately sample the neutrino phase space.  The simulations must
reflect the fact that the neutrinos initially have a redshifted
Fermi-Dirac phase space distribution \cite{KHPR}.  These CHDM
simulations were compared with observational data using various
statistics.  CHDM with $\Omega_\nu=0.3$, the value indicated by
approximate analyses \cite{HoltzPri,SShafi94}, was shown to
lead to groups of galaxies having substantially lower velocity
dispersions than CDM, and in better agreement with observations
\cite{NKP94}.  But it also leads to a Void Probability Function (VPF)
with more intermediate-sized voids than are observed \cite{Ghigna94}.
This theory had so little small-scale power that a quasi-linear
analysis using the Press-Schechter approximation
showed that there would not be enough of the high-column-density
hydrogen clouds at high redshift $z \sim 3$ known as damped
Lyman-$\alpha$ systems
\cite{KauffmannCharlot94,MoMiraldaEscude94,MaBert94}.  But CHDM with
$\Omega_\nu=0.2$ suppresses small-scale fluctuations less and therefore
has a better chance of avoiding this problem \cite{KBHP}.  Simulations
\cite{KNP} showed that this version of CHDM also has a VPF in good
agreement with observations \cite{Ghigna97}.  The group velocity
dispersions also remained sufficiently small to plausibly agree with
observations, but it had become clear that the N-body simulations used
lacked sufficient resolution to identify galaxies so that this
statistic could be measured reliably \cite{NKP98}.

A resolution problem also arose regarding the high-redshift damped
Lyman-$\alpha$ systems.  Earlier research had been based on the idea
that these systems are rather large disk galaxies in massive halos
\cite{MaBertetal97}, but then high-resolution hydrodynamical
simulations \cite{HaehneltSteinmetzRauch} showed that relatively small
gaseous protogalaxies moving in smaller halos provide a good match to
the new, detailed kinematic data \cite{ProchaskaWolfe}.  It thus
appeared possible that CHDM models with $\Omega_\nu \lsim 0.2$ might
produce enough damped Lyman-$\alpha$ systems.  With the low Hubble parameter
$h\sim0.5$ required for such $\Omega_m=1$ models, the total neutrino
mass would then be $\lsim 5$ eV.

While neutrino oscillation experiments can determine the differences of
squared neutrino masses, as we will briefly review next, cosmology is
sensitive to the actual values of the neutrino masses --- for any that
are larger than about 1 eV.  In that case, cosmology can help to fill in
the neutrino mass matrix.

One example of this is the fact that if the hot DM mass is roughly
evenly shared between two or three neutrino species, the neutrinos
will be lighter than if the same mass were all in one species, so that
the free streaming length will be longer.  A consequence is that, for
the same total neutrino mass and corresponding $\Omega_\nu$, the power
spectrum will be approximately 20\% lower on the scale of galaxy
clusters if the mass is shared between two neutrino species
\cite{Primack95}.  Since the amplitude and ``tilt'' $n$ of the power
spectrum in CDM-type models is usually fixed by comparison with COBE
and cluster abundance, this has the further consequence that higher
$n$ (i.e., less tilt) is required when the neutrino mass is divided
between comparable-mass neutrino species.  Less tilt means that there
will be more power on small scales, which appeared to be favorable for
the CHDM model, for example because it eased the problems with damped
Lyman-$\alpha$ systems \cite{Primack95,Pogosyan95}.

\section{Evidence for Neutrino Mass from Oscillations}

There is mounting astrophysical and laboratory data suggesting that
neutrinos oscillate from one species to another \cite{NuIndustry},
which can only happen if they have non-zero mass. Of these
experimental results, the ones that are regarded as probably most
secure are those concerning atmospheric neutrino oscillations from
Super-Kamiokande (see the chapter by John Learned) and solar neutrinos
from several experiments (see the chapter by Wick Haxton). But the
experimental results that are most relevant to neutrinos as hot dark
matter are from the Liquid Scintillator Neutrino Detector (LSND)
experiment at Los Alamos (see the chapter by David Caldwell).

Older Kamiokande data \cite{Kam} showed that, for events attributable
to atmospheric neutrinos with visible energy $E > 1.3$ GeV, the deficit
of $\nu_\mu$ increases with zenith angle.  The 
Super-Kamiokande detector has confirmed and extended the results of its
smaller predecessor \cite{SuperK}. These data imply that $\nu_\mu
\rightarrow \nu_\tau$ oscillations occur with a large mixing angle
$\sin^2 2\theta > 0.82$ and an oscillation length several times the
height of the atmosphere, which implies that $ 5 \times 10^{-4} <
\Delta m^2_{\tau \mu} < 6 \times 10^{-3}$ eV$^2$ (90\% CL).  (Neutrino
oscillation experiments measure not the masses, but rather the
difference of the squared masses, of the oscillating species, here
$\Delta m_{\tau \mu}^2 \equiv |m(\nu_\tau)^2 - m(\nu_\mu)^2|$.) This in
turn implies that if other data requires either $\nu_\mu$ or $\nu_\tau$
to have large enough mass ($\gsim 1$ eV) to be a hot dark matter
particle, then they must be nearly equal in mass, i.e., the hot dark
matter mass would be shared between these two neutrino species.  Both
the new Super-Kamiokande atmospheric $\nu_e$ data and the lack of a
deficit of $\bar\nu_e$ in the CHOOZ reactor experiment \cite{chooz}
make it quite unlikely that the atmospheric neutrino oscillation is
$\nu_\mu \rightarrow \nu_e$.  If the oscillation were instead to a
sterile neutrino, the large mixing angle implies that this sterile
species would become populated in the early universe and lead to too
much $^4$He production during the Big Bang Nucleosynthesis epoch
\cite{shi}.  (Sterile neutrinos are discussed further below.) It may be
possible to verify that $\nu_\mu \rightarrow \nu_\tau$ oscillations
occur via a long-baseline neutrino oscillation experiment. The K2K
experiment is looking for missing $\nu_\mu$ due to $\nu_\mu \rightarrow
\nu_\tau$ oscillations with a beam of $\nu_\mu$ from the Japanese KEK
accelerator directed at the Super-Kamiokande detector, with more
powerful Fermilab-Soudan and CERN-Gran Sasso long-baseline experiments
in preparation, the latter of which will look for $\tau$ appearance.

The observation by LSND of events that appear to represent $\bar\nu_\mu
\rightarrow \bar\nu_e$ oscillations followed by $\bar \nu_e + p \to n +
e^+$, $n + p \to D + \gamma$, with coincident detection of $e^+$ and
the 2.2 MeV neutron-capture $\gamma$-ray, suggests that $\Delta m_{\mu
e}^2 > 0$ \cite{lsndan}. The independent LSND data \cite{lsndn}
suggesting that $\nu_\mu \rightarrow \nu_e$ oscillations are also
occurring is consistent with, but has less statistical weight than, the
LSND signal for $\bar\nu_\mu \rightarrow \bar\nu_e$ oscillations.
Comparison of the latter with exclusion plots from other experiments
allows two discrete values of $\Delta m^2_{\mu e}$, around 10.5 and 5.5
eV$^2$, or a range 2 eV$^2 \gsim \Delta m^2_{\mu e} \gsim 0.2$ eV$^2$.
The lower limit in turn implies a lower limit $m_\nu \gsim 0.5$ eV, or
$\Omega_\nu \gsim 0.01 (0.65/h)^2$.  This would imply that the
contribution of hot dark matter to the cosmological density is at least
as great as that of all the visible stars $\Omega_\ast \approx 0.0045
(0.65/h)$ \cite{FukugitaHP}.  Such an important conclusion requires
independent confirmation.  The KArlsruhe Rutherford Medium Energy
Neutrino (KARMEN) experiment has added shielding to decrease its
background so that it can probe a similar region of $\Delta m^2_{\mu
e}$ and neutrino mixing angle; the KARMEN results exclude a significant
portion of the LSND parameter space, and the numbers quoted above take
into account the current KARMEN limits.  The Booster Neutrino
Experiment (BOONE) at Fermilab should attain greater sensitivity.

The observed deficit of solar electron neutrinos in three different
types of experiments suggests that some of the $\nu_e$ undergo
Mikheyev-Smirnov-Wolfenstein matter-enhanced oscillations $\nu_e
\rightarrow \nu_x$ to another species of neutrino $\nu_x$ with
$\Delta m_{e x}^2 \approx 10^{-5}$ eV$^2$ as they travel through the
sun \cite{solarnu},
or possibly ``Just-So'' vacuum oscillations with even smaller
$\Delta m_{e x}^2$ \cite{justso}.
The LSND $\nu_\mu \rightarrow \nu_e$ signal with a much larger $\Delta
m_{e \mu}^2$ is inconsistent with $x=\mu$, and the Super-Kamiokande
atmospheric neutrino oscillation data is inconsistent with $x=\tau$.  Thus
a fourth neutrino species $\nu_s$ is required if all these neutrino
oscillations are actually occurring.  Since the neutral weak boson
$Z^0$ decays only to three species of neutrinos, any additional
neutrino species $\nu_s$ could not couple to the $Z^0$, and is called
``sterile.''  This is perhaps distasteful, although many modern
theories of particle physics beyond the standard model include the
possibility of such sterile neutrinos.  The resulting pattern of
neutrino masses would have $\nu_e$ and $\nu_s$ very light, and
$m(\nu_\mu) \approx m(\nu_\tau) \approx (\Delta m_{e \mu}^2)^{1/2}$,
with the $\nu_\mu$ and $\nu_\tau$ playing the role of the hot dark
matter particles if their masses are high enough \cite{fournu}.  This
neutrino spectrum might also explain how heavy elements are synthesized
in core-collapse supernova explosions \cite{caldrev}.
Note that the required solar neutrino mixing angle is very small,
unlike that required to explain the atmospheric $\nu_\mu$ deficit, so a
sterile neutrino species would not be populated in the early universe and
would not lead to too much $^4$He production.

Of course, if one or more of the indications of neutrino oscillations
are wrong, then a sterile neutrino would not be needed and other
patterns of neutrino masses are possible.  But in any case the
possibility remains of neutrinos having large enough mass to be hot dark
matter.  Assuming that the Super-Kamiokande data on atmospheric
neutrinos are really telling us that $\nu_\mu$ oscillates to $\nu_\tau$,
the two simplest possibilities regarding neutrino masses are as follows:

\noindent {\bf A}) Neutrino masses are hierarchical like all the other
fermion masses, increasing with generation, as in see-saw models.  Then
the Super-Kamiokande $\Delta m^2 \approx 0.003$ implies $m(\nu_\tau)
\approx 0.05$ eV, corresponding to 
\begin{equation}
\Omega_\nu = 0.0013 (m_\nu/0.05 {\rm eV}) (0.65/h)^2 \ .
\end{equation}
This is not big enough to affect galaxy
formation significantly, but it is another puzzling cosmic coincidence
that it is close to the contribution to the cosmic density from stars.

\noindent {\bf B}) The strong mixing between the mu and tau neutrinos
implied by the Super-Kamiokande data suggests that these neutrinos are
also nearly equal in mass, as in the Zee model \cite{Zee} and many
modern models \cite{justso,fournu} (although such strong mixing can
also be explained in the context of hierarchical models based on the
SO(10) Grand Unified Theory \cite{BabuPW}).  Then the above
$\Omega_\nu$ is just a lower limit.  An upper limit is given by
cosmological structure formation.  In Cold + Hot Dark Matter (CHDM)
models with $\Omega_m=1$, we saw in the previous section that if
$\Omega_\nu$ is greater than about 0.2 the voids are too big and there
is not enough early structure.  In the next section we consider the
upper limit on $\Omega_\nu$ if $\Omega_m \approx 0.4$, which is favored
by a great deal of data.

\section{Cold plus Hot Dark Matter and Structure Formation: $\Omega_m
\approx 0.4$}

We have already mentioned that the $\Omega_m=1$ CHDM model with
$\Omega_\nu=0.2$ was found to be the best fit to nearby galaxy data of
all cosmological models \cite{GawiserSilk98}. But this didn't take
into account the new high-z supernova data and analyses \cite{hizsn}
leading to the conclusion that $\Omega_\Lambda - \Omega_{matter}
\approx 0.2$, nor the new high-redshift galaxy data. Concerning the
latter, Somerville, Primack, and Faber \cite{spf} found that none of
the $\Omega_m=1$ models with a realistic power spectrum (e.g., CHDM,
tilted CDM, or $\tau$CDM) makes anywhere near enough bright $z \sim 3$
galaxies.  But we found that \lcdm\ with $\Omega_m \approx 0.4$ makes
about as many high-redshift galaxies as are observed \cite{spf}.  This
$\Omega_m$ value is also implied if clusters have the same baryon
fraction as the universe as a whole: $\Omega_m \approx \Omega_b / f_b
\approx 0.4$, using for the cosmological density of ordinary matter
$\Omega_b = 0.019 h^{-2}$ \cite{BurlesTytler98} and for the cluster
baryon fraction $f_b = 0.06 h^{-3/2}$ \cite{Evrard97} from X-ray data
or $f_b = 0.077 h^{-1}$ from Sunyaev-Zel'dovich data
\cite{Carlstrom}. An analysis of the cluster abundance as a function
of redshift based on X-ray temperature data also implies that $\Omega_m
\approx 0.44\pm0.12$ \cite{ecfh,Henry00}. Thus most probably $\Omega_m$ is
$\sim 0.4$ and there is a cosmological constant $\Omega_\Lambda \sim
0.6$.  In the 1984 paper that helped launch CDM \cite{BFPR}, we
actually considered two models in parallel, CDM with $\Omega_m=1$ and
\lcdm\ with $\Omega_m=0.2$ and $\Omega_\Lambda=0.8$, which we thought
would bracket the possibilities.  It looks like an \lcdm\ intermediate
between these extremes may turn out to be the right mix.

\begin{figure}
\centering
\centerline{\psfig{file=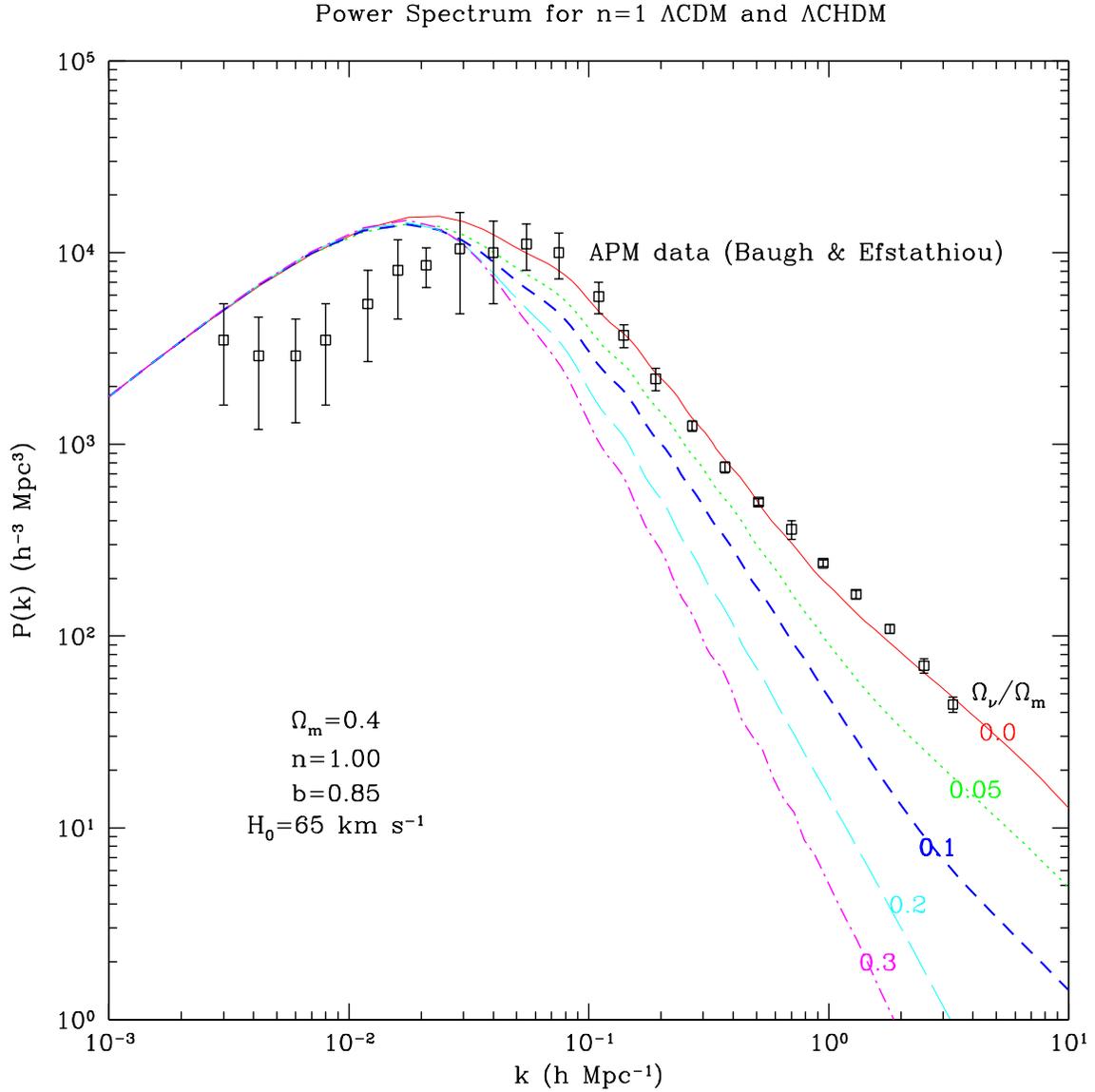,height=16cm}}
\caption{
Nonlinear dark matter power spectrum vs. wavenumber for \lcdm\ and
\lchdm\ models with $\Omega_\nu/\Omega_m=0.05, 0.1, 0.2, 0.3$.  Here
$\Omega_m=0.4$, the Hubble parameter $h=0.65$, there is no tilt (i.e.,
$n=1$), and the bias $b=0.85$.
Note that in this and the next Figure we
``nonlinearized'' all the model power spectra \protect\cite{Smith98}, to
allow them all to be compared to the APM data (the small ``wiggles'' in the
high-$\Omega_\nu$ power spectra are an artifact of the nonlinearization
procedure).  
}
\end{figure}

\begin{figure}
\centering
\centerline{\psfig{file=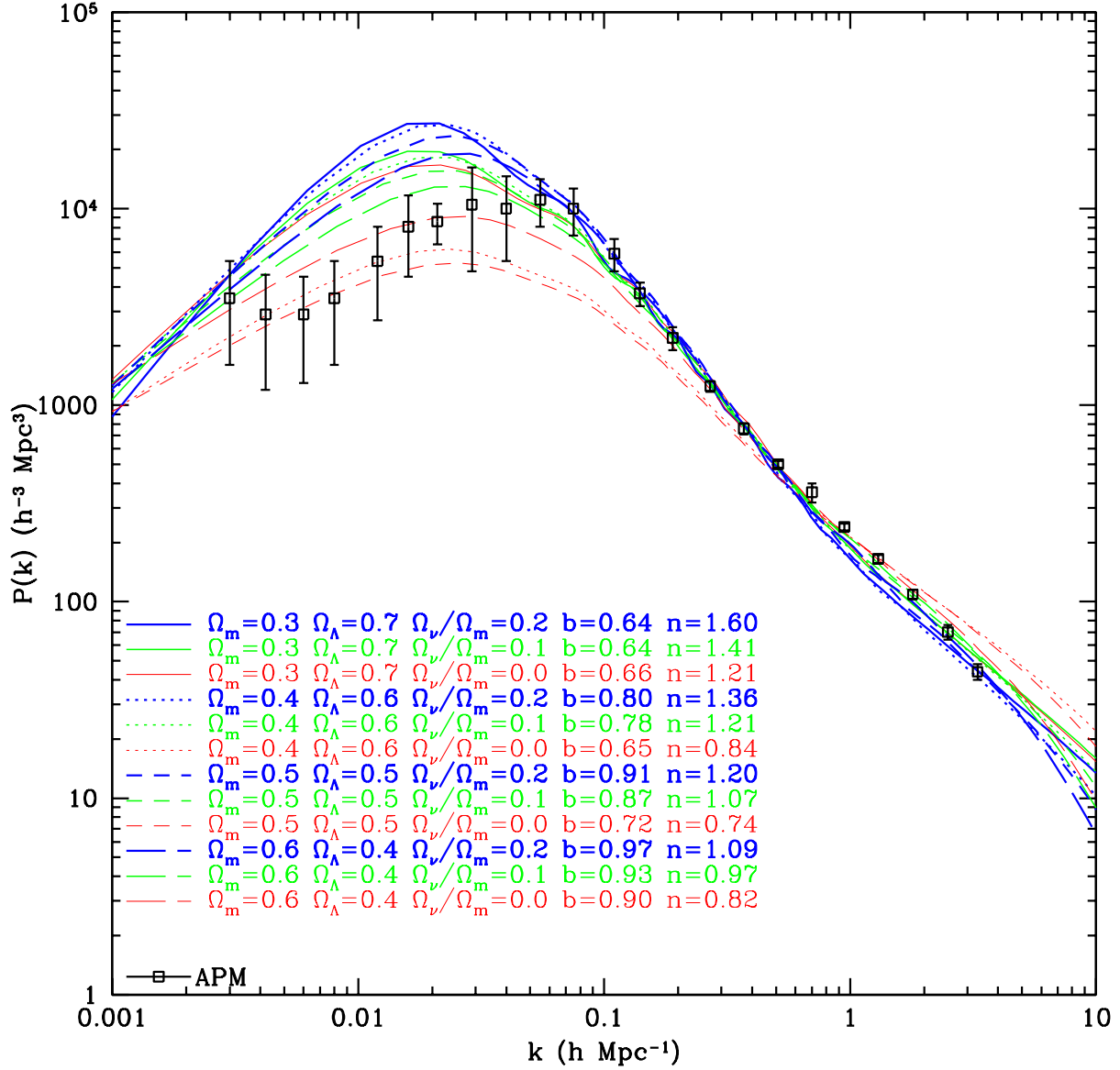,height=17cm}}
\caption{ 
Nonlinear dark matter power spectrum vs.  wavenumber for 12
\lchdm\ models with $N_\nu=2$ massive neutrino species and Hubble
parameter $h=0.65$, with tilt and $\sigma_8$ determined by COBE +
ENACS cluster abundance.  The bias chosen for these models is that
which minimizes $\chi^2$ over the entire range of available APM data.
}
\end{figure}

\begin{figure}
\centering
\centerline{\psfig{file=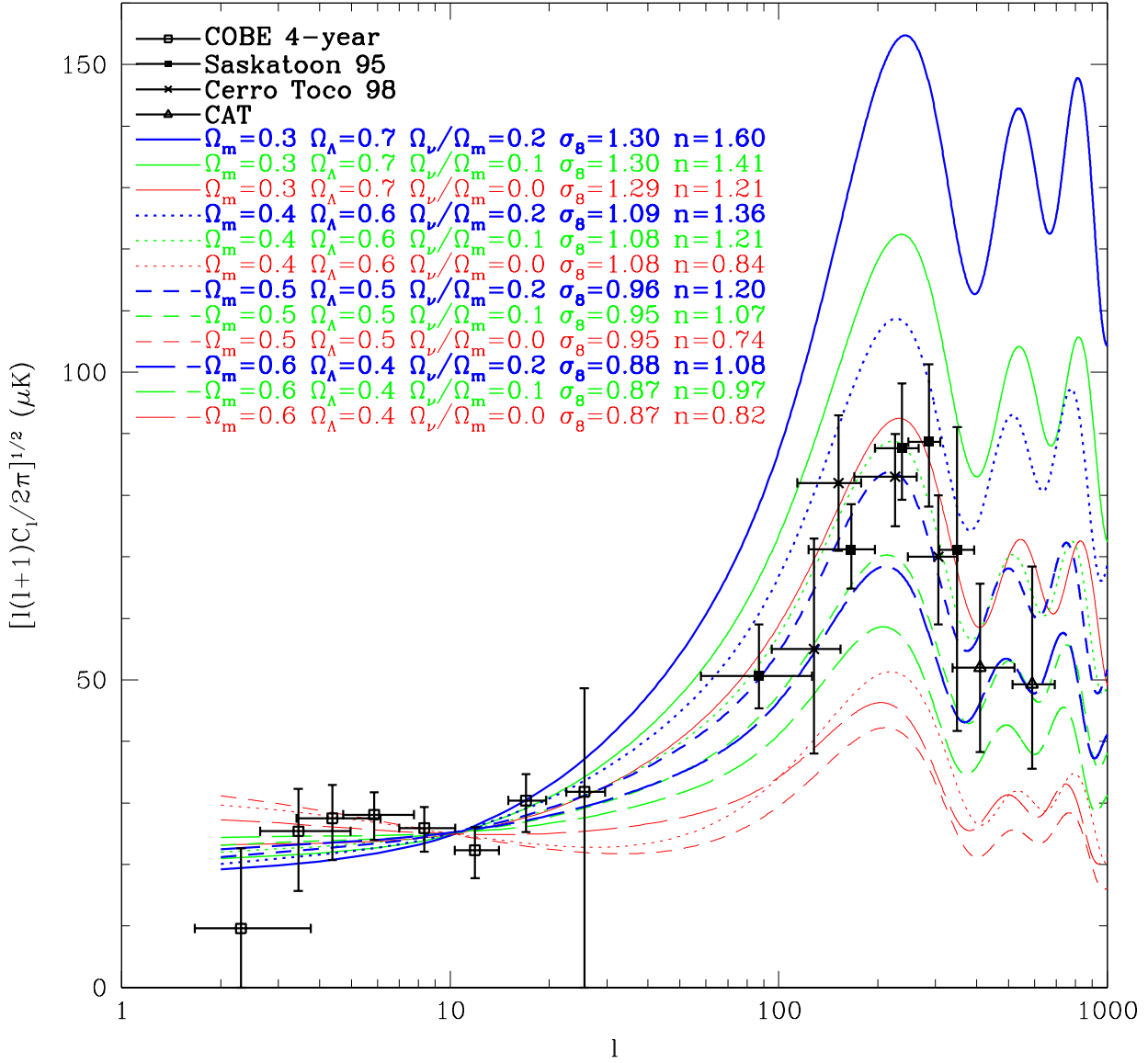,height=17cm}}
\caption{CMB anisotropy power spectrum vs. angular wave
number for the same models as in Figure 2.
The data plotted are from COBE and three recent small-angle experiments
\protect\cite{sk,cat1,cat2,toco}.
}
\end{figure}

The success of $\Omega_m=1$ CHDM in fitting the CMB and galaxy
distribution data suggests that flat low-$\Omega_m$ cosmologies with a
little hot dark matter be investigated in more detail.  We
have used CMBFAST \cite{Seljak} to examine \lchdm\ models with various
$h$, $\Omega_m$, and $\Omega_\nu$, assuming $\Omega_b = 0.019 h^{-2}$.
Figure 1 shows the power spectrum $P(k)$ for \lcdm\ and a sequence 
of \lchdm\ models with increasing amounts of hot dark matter, compared
to the power spectrum from APM \cite{apm}.  Here we have fixed 
$\Omega_m=0.4$ and Hubble parameter $h=0.65$.  All
of these models have no tilt and the same bias parameter, to make it
easier to compare them with each other.  As expected, the large-scale
power spectrum is the same for all these models, but the amount
of small-scale power decreases as the amount of hot dark matter increases.

In Figures 2 and 3 we consider a sequence of twelve \lcdm\ and \lchdm\
models with $h=0.65$, $\Omega_m=0.3,0.4,0.5,0.6$, and 
$\Omega_\nu/\Omega_m=0,0.1,$ and 0.2.
We have adjusted the amplitude and tilt $n$ of the primordial
power spectrum for each model in order to match the 4-year COBE amplitude
and the ENACS differential mass function of clusters \cite{ENACS}
(cf. \cite{Grossprep}).  (We checked the
CMBFAST calculation of \lchdm\ models against Holtzman's code used in
our earlier investigation of \lchdm\ models \cite{Primack95}.  Our
results are also compatible with those of recent studies
\cite{Valdarnini,Fukugita00} in which $n=1$ models were considered.  But we
find that some \lcdm\ and
\lchdm\ models require $n>1$, called ``anti-tilt'', and
it is easy to create cosmic inflation models that give $n>1$ ---
cf. \cite{Bono98}.)  In all the \lchdm\ models the neutrino mass is
shared between $N_\nu=2$ equal-mass species --- as explained above, this
is required by the atmospheric neutrino oscillation data if neutrinos
are massive enough to be cosmologically significant hot dark matter.  
(This results in slightly
more small-scale power compared to $N_\nu=1$ massive
species, as explained above, but the $N_\nu=1$ curves are very similar to
those shown.)
In Ref. \cite{PriGrossBlois} we have shown similar results for Hubble
parameter $h=0.6$, and also plotted the best CHDM and $\Lambda$CDM 
models from \cite{GawiserSilk98}.
Note that all these Figures are easier to read in color; see the version
of this paper on the Los Alamos archive \cite{PriGross00}.

\begin{figure}
\centering
\centerline{\psfig{file=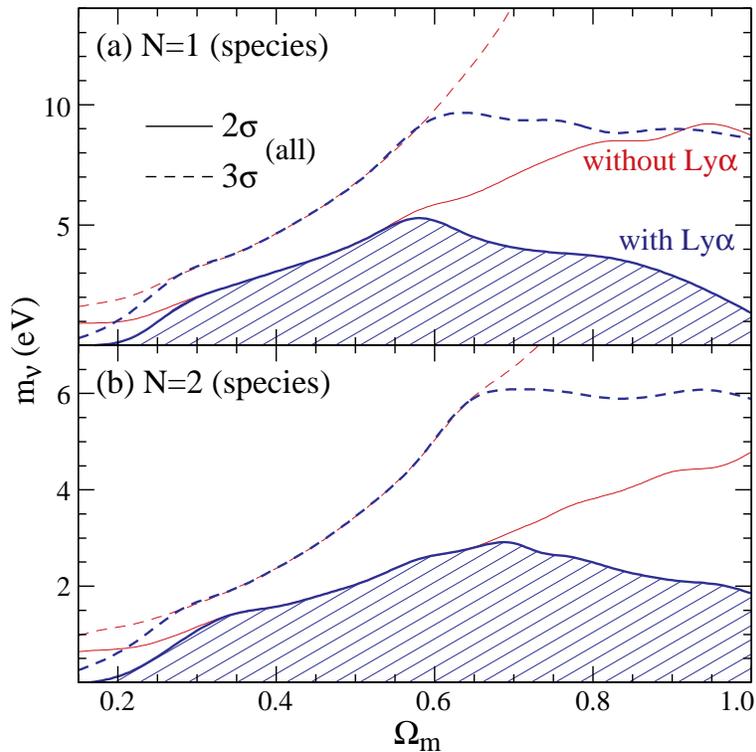,height=10cm}}
\caption{
Constraints on the neutrino mass assuming (a) $N_\nu=1$ massive
neutrino species and (b) $N_\nu=2$ equal-mass 
neutrino species.  The heavier weight curves show the effect 
of including the Lyman-$alpha$ forest constraint.  (From 
\protect\cite{CroftHD}; used by permission.)
}
\end{figure}

Of the \lchdm\ models shown, for $\Omega_m=0.4 - 0.6$ the best
simultaneous fits to the small-angle CMB and the APM galaxy power
spectrum data \cite{apm} are obtained for the model with
$\Omega_\nu/\Omega_m=0.1$, and correspondingly $m(\nu_\mu) \approx
m(\nu_\tau) \approx 0.8 - 1.2$ eV for $h=0.65$.
For $\Omega_m < 0.4$, smaller or vanishing neutrino mass appears
to be favored.
Note that the anti-tilt permits some of the \lchdm\
models to give a reasonably good fit to the COBE plus small-angle CMB data.
Thus, adding a little hot dark matter to
the moderate-$\Omega_m$ \lcdm\ models may perhaps 
improve somewhat their simultaneous
fit to the CMB and galaxy data, but the improvement is not nearly as
dramatic as was the case for $\Omega_m=1$.

It is apparent that the \lcdm\ models with $\Omega_m=0.4,0.5$ have too
much power at small scales ($k \gsim 1\hmpc$), as is well known
\cite{KPH,Jenkins} --- although recent work \cite{Colin} suggests that
the distribution of {\it dark matter halos} in the $\Omega_m=0.3$,
$h=0.7$ \lcdm\ model may agree well with the APM data.  On the other
hand, the \lchdm\ models may have too little power on small scales ---
high-resolution \lchdm\ simulations and semi-analytic models of early
galaxy formation may be able to clarify this.  Such simulations should
also be compared to data from the massive new galaxy redshift surveys
2dF and SDSS using shape statistics, which have been shown to be
able to discriminate between CDM and CHDM models \cite{Dave}.

Note that all the \lcdm\ and \lchdm\ models that are normalized to
COBE and have tilt compatible with the cluster abundance are a poor
fit to the APM power spectrum near the peak.  The \lchdm\ models all
have the peak in their linear power spectrum $P(k)$ higher and at lower
$k$ than the currently available data (e.g., from APM).  Thus the
viability of \lcdm\ or \lchdm\ models with a power-law primordial
fluctuation spectrum (i.e., just tilt $n$) depends on this
data/analysis being wrong.  In fact, it has recently been argued
\cite{Eisenstein} that because of correlations, the errorbars
underestimate the true errors in $P(k)$ for small $k$ by at least a
factor of 2.  The new large-scale surveys 2dF and SDSS will be crucial
in giving the first really reliable data on this, perhaps as early as
next year.

The best published constraint on $\Omega_\nu$ in \lchdm\ models is
\cite{CroftHD}.  Figure 4 shows the result of their analysis, which
uses the COBE and cluster data much as we did above, the $P(k)$ data
only for $0.025 (h/{\rm Mpc}) < k < 0.25 (h/{\rm Mpc})$, the
constraint that the age of the universe is at least $13.2\pm2.9$ Gyr
(95\% C.L.)  from globular clusters \cite{Carretta}, and also the
power spectrum at high redshift $z\sim 2.5$ determined from
Lyman-$\alpha$ forest data.  The conclusion is that the total neutrino
mass $m_\nu$ is less than about 5.5 eV for all values of $\Omega_m$,
and $m_\nu \lsim 2.4(\Omega_m/0.17-1)$ eV for the observationally
favored range $0.2 \leq \Omega_m \leq 0.5$ (both at 95\% C.L.).
Analysis of additional Lyman-$\alpha$ forest data can allow detection
of the signature of massive neutrinos even if $m_\nu$ is only a
fraction of an eV.  Useful constraints on $\Omega_\nu$ will also come
from large-scale weak gravitational lensing data \cite{Cooray}
combined with cosmic microwave background anisotropy data.

JRP acknowledges support from NASA and NSF grants at UCSC.


%
\vfill
\end{document}